**AI and the future of pharmaceutical research**

Adam Zielinski

Thomas Edison State University

**Abstract**

This paper examines how pharmaceutical Artificial Intelligence (AI) advancements may affect the development of new drugs in the coming years. The question was answered by reviewing a rich body of source material, including industry literature, research journals, AI studies, market reports, market projections, discussion papers, press releases, and organizations' websites. The paper argues that continued innovation in pharmaceutical AI will enable rapid development of safe and effective therapies for previously untreatable diseases. A series of major points support this conclusion: The pharmaceutical industry is in a significant productivity crisis today, and AI-enabled research methods can be directly applied to reduce the time and cost of drug discovery projects. The industry already reported results such as a 10-fold reduction in drug molecule discovery times. Numerous AI alliances between industry, governments, and academia enabled utilizing proprietary data and led to outcomes such as the largest molecule toxicity database to date or more than 200 drug safety predictive models. The momentum was recently increased by the involvement of tech giants combined with record rounds of funding. The long-term effects will range from safer and more effective therapies, through the diminished role of pharmaceutical patents, to large–scale collaboration and new business strategies oriented around currently untreatable diseases. The paper notes that while many reviewed resources seem to have overly optimistic future expectations, even a fraction of these developments would alleviate the productivity crisis. Finally, the paper concludes that the focus on pharmaceutical AI put the industry on a trajectory towards another significant disruption: open data sharing and collaboration.

*Keywords:* Drug research, Artificial Intelligence, Pharmaceutical AI, Drugs, Machine learning, Medicines, Productivity crisis.

## Table of contents

**Introduction**

Artificial Intelligence (AI) is a relatively new field, yet it has already enabled inventions ranging from Netflix recommendations, through voice assistants, to self-driving cars. AI could also revolutionize pharmaceutical research and development (R&D), the productivity of which has been steadily decreasing and reached the lowest levels in history (Paul et al., 2010; Paul et al., 2020; Pammolli et al., 2011). Several recent studies reported how AI methods might streamline drug research by enabling faster identification of drug targets (Bravo et al., 2015), simplifying the design of drug molecules (Zhavoronkov et al., 2019a), and improving the design of preclinical trials (Pound, 2019). Besides, numerous hypothesized advancements may enable entirely new types of treatments and even revolutionize the entire pharmaceutical industry. While both the early results and the future projections are promising, many proposed methods are purely theoretical, while other suffer from deficiencies and need to be further refined before being widely adopted. Today, the pharmaceutical AI is still in a young research branch, posing more questions than answers.

The field concerned with advancing AI-based methods is called *Data Science*. One of the challenges in data science today is the available computational power. Solving specific drug research problems using massive amounts of medical data would take years on today's machines (Chen et al., 2020)**.** Fortunately, according to Moore's law, the available computational power doubles every two years (Encyclopaedia Britannica, 2019). A large supply of experts combined with faster computers will help develop effective AI-enabled drug research techniques and lead to rapid discovery of affordable drugs for previously untreatable diseases.

# Literature review

## Artificial Intelligence is merely applied statistics

The name *Artificial Intelligence* may be misleading. Popular culture often portrays AI as a type of sentient machine capable of reasoning about the world on its own, in some instances even being more intelligent than any human. Movies like *Terminator* went one step further and introduced the idea of a life-threatening super AI. In reality, today's AI is merely a practical application of statistical methods via computer software (Smith & Eckroth, 2017; Encyclopaedia Britannica, 2020). AI can be broadly divided into machine learning (ML), data mining, optimization, and statistics.

What distinguishes AI-based software from regular software is the way in which it solves a problem. Classical computer programs consist of predefined instructions exhaustively describing a program's behavior; for example, an accounting software would likely contain exact recipes for generating yearly statements, rounding the numbers, and calculating tax. Such an approach makes it easy to answer specific questions like *how much tax do I owe?* but cannot be easily applied to more open-ended problems. For example, it is not trivial to specify a set of rules for deciding whether a photo depicts a car or a bird. Such problems are precisely where AI-based software shines: Instead of using a human-preconceived notion of cars and birds, an AI program would analyze a large set of photos to identify patterns distinguishing ones from the others – a procedure known as *training*. Once a model is trained, it may be used to look for the identified patterns in previously unseen photos to classify them as either cars or birds. The computer-identified patterns are often not easily interpretable by humans, which brings an interesting conclusion: While the solution to problems like image classification is too complicated for humans to formalize directly, scientists found a way of inferring answers from relevant data.

The applications of AI go beyond image classification. For example, a more advanced AI model could generate new photos based on the training examples it was exposed to (OpenAI, 2020). As this project will show later, an expansion of that idea led to generating accurate predictions of drug targets' 3D structures (Deloitte, 2019b). All in all – the AI-based approach enables humans to reason about vast amounts of data and convert them into solutions to problems too hard for humans to solve directly.

**Prevalence of productivity crisis in pharmaceutical R&D**

The process of discovering drugs is long and expensive. The estimated cost and time required to introduce a new drug to the market varies from US $500 billion and 9 years to US $2.6 billion and 13.9 years (Adams & Brantner, 2006; Cockburn, 2006; Kola & Landis, 2004; Pammolli et al., 2011; Scannel et al., 2012). A recent study analyzed the industry and government data to determine the number of approved drugs per the inflation adjusted US $1 billion investment (Scannell et al., 2012). It concluded that "the number of new drugs approved per billion US dollars spent on R&D has halved roughly every 9 years since 1950, falling around 80−fold in inflation−adjusted terms" (Scannell et al., 2012). Similar findings were reported in several independent studies, legitimizing the issue (Kola & Landis, 2004; Pammolli et al., 2011; Paul et al., 2010).

The industry also suffers from a high percentage of failed candidate drugs, also known as *attrition rates*. For every 5,000 - 10,000 investigated drug compounds, only ten complete preclinical trials (Kola & Landis, 2004), and only one is approved by Food and Drug Administration (FDA) (Deore et al., 2019; Torjesen, 2019). The attrition rates of candidate drugs in clinical trials have been fluctuating over the years but remained mostly constant since 1990 (DiMasi et al., 2010; Kola & Landis, 2004; Wong et al., 2018).

Not all researchers agree with these conclusions. Cockburn (2006) raised two counter-points: first, that reported R&D expenditures are frequently overstated as they lack adjustment for inflation – an argument contradicted in the inflation-adjusted study by Scannell et al. (2012). Second, increased investments only bring results after ten years. Cockburn (2006) predicted an increase in approved drugs in 2016 – 10 years after the study. Indeed, the number of FDA drug approvals increased from 209 between 2000 and 2008 to 302 between 2009 and 2017 (Batta, 2020). Still, the average research cost doubled again from the US $1.188 billion in 2010 to the US $2.168 billion in 2018 (Deloitte, 2018). Another study analyzed the dataset of 50,000 drug development projects and concluded the attrition rates have actually decreased between 1990 and 2013 (Pammolli et al., 2020). The numeric results are consistent with those other studies, but the conclusions drawn are vastly different as the researchers questioned the productivity crisis's existence. In contrast, other studies noted that attrition rates fluctuate over time and that the productivity crisis must be considered with other variables in mind.

All in all, the increase in time and cost combined with the flat attrition rate dramatically decreased the return on drug development investments: From 10.1% in 2010 to 1.8% in 2019 (Deloitte, 2019b). The continuation of these unsettling trends would delay the development of drugs for untreatable diseases and possibly cause a collapse of drug innovation. Pharmaceutical research and development need a pivotal change to streamline the development of new drugs.

**The drug development process**

In general, drugs work by interacting with *targets* – proteins, genes, and nucleic acids relevant to the disease (Deore, 2019). A famous analogy represents this process in terms of a lock (target) and a key (drug). Developing a new drug requires finding the right lock, designing a key that would fit into it, and conducting clinical trials to make sure if the key indeed fits into the lock. A good key should also not accidentally open any other locks (side-effects).

The drug development process may be broadly split into two sequential stages: research and clinical trials. The research stage takes an average of 5 to 6 years. Most of that time is spent processing data and performing simulations (Deore, 2019), making Artificial Intelligence an excellent tool to expedite the process. Clinical trials last for 5 to 7 years on average, are focused on verifying the efficacy and safety of proposed drugs (Deore, 2019).

**Artificial intelligence applications at the research stage**

***Target identification***

Drug research starts with *target identification*, a process of reviewing the literature and analyzing target databases in search of targets playing a role in the disease (Deore, 2019). The availability of relevant prior research, typically conducted at academia and other research centers, is critical to this step's success (Matthews et al., 2016). Target identification is also the most critical part of the process. It determines the entire research project's success – failing to select a treatable target will ensure a failure years later. Recent studies have shown how machine learning may be used to accelerate the identification of treatable targets. Costa et al. (2010) proposed a classification method that categorizes many potential targets as treatable/non-treatable based on prior knowledge about the properties of already identified targets. Two other studies described how *Natural Language*

*Processing* (NLP) techniques were used for automatic extraction of the relationship between targets and diseases from the MEDLINE database containing a large body of research (Bravo et al., 2015; Kim et al., 2017). It is worth noting that future improvements of such NLP techniques could foster a scientific debate by enabling quick identification of studies reporting contradictory results.

### Target validation

In the next step, *target validation,* the identified candidate targets are narrowed down to the most promising ones (Deore, 2019). A good target exhibits two essential characteristics. First, it plays an essential role in the disease. Second, it has surface cavities of such shape and size that a potential drug could bind with it (Deore, 2019). Several recent studies proposed automating different parts of target validation: Google developed a machine learning algorithm that predicts the structure of identified targets more accurately than experienced field experts (Hutson, 2019). Besides, Nayal and Honig proposed a classifier trained using data about known protein cavities to making predictions about whether supposed targets are "druggable" or not (2006).

### Lead identification

Once a promising target is identified, the next step is *lead identification*. Lead is a chemical molecule that is likely to bind with a specific target while exhibiting several properties such as low toxicity (Deore, 2019). In a recent study, Zhavoronkov et al. (2019a) described a machine learning method of lead identification that allowed them to find a candidate fibrosis drug in a matter of weeks compared to months or years. Similarly, Imperial College identified 110 candidate anti-cancer molecules in an innovative study where several volunteers agreed to run AI-based computations using their phones' idle times (Veselkov, 2019).

*Lead optimization*

Drug molecules may interact not just with the intended target but also with several other proteins causing unwanted side effects. In the next step, l*ead optimization*, candidate molecules are evaluated and refined. The goal is to reduce the chance of side effects and improve several properties such as stability, specificity, or toxicity. Farimani et al. (2018) proposed a machine learning technique to predict lead molecules' properties and other similar compounds using significantly less time and data compared to traditional statistical methods.

*Preclinical trials*

Once the target is identified and several lead compounds are selected, their efficacy and safety in verified in trials on animals, also known as *preclinical research* or *preclinical trials*. One of the most considerable problems of preclinical research is that drug molecules may exhibit different behavior in animals than in humans (Pound & Ritskes-Hoitinga, 2018). To remedy the problem, Normand et al. proposed a new ML method to predict cross-species differences between mice and humans (2019). The information identified by the model could help identify false leads early in the process and prevent bearing the cost of clinical trials.

**Alternative drug research and development methods**

The ability to efficiently process large amounts of data led to the emergence of novel, computer-aided drug research methods. Such innovations include repurposing existing drugs for other diseases (Zhavoronkov et al., 2019b), personalized drugs (Schneider et al., 2020), or analyzing cancer tissue to directly identify effective chemical compounds (Narain et al., 2011; Reiss, 2020). These methods deserve a proper introduction and will be explored in more detail in the final project. This document only discusses a single novel method called *de novo* drug design, which is discussed in several studies, making it a perfect fit for the literature review.

### De novo drug design

A *de novo* design method could streamline lead identification and optimization, enabling rapid discovery of therapeutically useful molecules. Instead of screening known molecules for desired properties, *de novo* design uses ML to predict novel drug molecules exhibiting desired properties based on the existing knowledge (Schneider & Clark, 2020). Using a lock and key analogy, instead of using a key that already exists, *de novo* design crafts a perfect key from scratch. The idea received significant attention during the last decade, and many studies proposed an abundance of *de novo* design techniques (Lin et al., 2020). De novo design was recently applied by Zhavoronkov et al. (2019a) to generate and validate the lead molecule for fibrosis treatment.

While many papers are published about *de* novo design (Lin et al., 2020), Schneider and Clark (2020) pointed out that only a few instances of practical applications are known. Several unresolved problems still pose a challenge, preventing wider adoption of the method. One such problem is the diversity of generated molecules: Because most candidate drugs fail, it is essential to test a diverse set of lead molecules to maximize chances of success. Nevertheless, state-of-the-art methods tend to generate many similar molecules (Benheda, 2017). Another challenge is that de novo design is computationally demanding and may find more applications when more computational power is available in the future (Chen et al., 2020)**.** Overcoming the challenges posed by *de novo* design would significantly reduce the time required to develop new drugs and, in turn, revolutionize the industry.

### Personalized therapies

An important research direction in pharmaceutical AI is the personalization of drug therapies. Depending on the degree of advancement of future AI-based techniques, there are several possible outcomes.

In the most basic scenario, the AI could enable medical professionals to identify the most promising treatment for specific patients. The studies on automated literature review like Bravo et al. (2015) and Kim et al. (2017) are a step in this direction. In addition, Microsoft launched *Project Hanover*, a platform capable of automatically reviewing literature in search of effective cancer treatments given patients' specific profile (Microsoft, 2019). FDA's INFORMED initiative set an even more ambitious goal of an industry-wide "transformation from a reductionist approach to drug development (for example, a single drug targeting a driver mutation and traditional clinical trials) to a holistic approach (for example, combination therapies targeting complex multiomic signatures and real-world evidence)" (Khozin, Kim, & Pazdur, 2017). AI-driven drug personalization efforts are expected to receive even more attention in the coming years (KPMG, 2020).

In a slightly more optimistic scenario, the pharmaceutical companies will start developing medicines targeted at specific patient populations. One possible research avenue is gene therapies (Zhavoronkov et al., 2019b). Such treatments already exist today and received significant focus in recent years – especially after the FDA approved the first medication of this type called *Kymriah* in 2017 (FDA, 2017). Since then, 18 more gene therapy products were approved, and numerous industry guidelines were released in anticipation of more gene–based medications (FDA, 2021; n.d.). Despite the large wave of approved gene therapies, their development remains an inherently challenging task due to a large number of variables and possible side–effects that must be considered (Zhavoronkov et al., 2019b). This point is best illustrated by the Kymriah medicine overview by European Medicines Agency (n.d.):

> Serious side effects occur in most patients. The most common serious side effects are cytokine release syndrome (a potentially life-threatening condition that can cause fever, vomiting, shortness of breath, pain, and low blood pressure). (…) Serious infections occur in around 3 in 10 DLBCL patients.

Turning to AI–based techniques could help conquer these challenges, especially considering the growing availability of genomics data (Zhavoronkov et al., 2019b). Even today, several AI companies and tech giants are working on combining the data and technology to enable reliable development of future gene-therapies (Deloitte, 2019a; KPMG, 2020). One such hypothesized therapy is called *Telomerase gene therapy* (Boccardi & Herbig, 2012). *Telomeres*, the sequences of molecules at the end of human chromosomes, protect our genetic material from damage (Boccardi & Herbig, 2012). The length of telomeres has recently been linked with longevity – the shorter the telomeres, the more severe are the effects of aging. The causes of telomeres shortening are not well understood at this point: some believe oxidative stress plays an essential role in this process, while others present experimental results to the contrary (Armanios et al., 2009; Boonekamp, 2017). There are no known ways of lengthening the telomeres; hence Boccardi and Herbig presented the idea of telomerase gene therapy, which would increase the length of telomeres with genetic manipulation (2012). Such a treatment could potentially slow aging and remedy age-related diseases.

Finally, in the most optimistic scenario, AI-based software would enable computational (*in silico*) simulations of candidate drug molecules' behavior and parameters (Deloitte, 2019b; Schneider et al., 2020). This research direction already received significant attention in recent years, which led to the development of predictive models of drug molecule safety profiles and toxicity levels (eTOX Project, 2010; MELLODDY Project, 2019). In the future, advancing these techniques even further could bring powerful research tools capable of aiding or even partially replacing clinical trials (Deloitte, 2019b; Zhavoronkov et al., 2019b). Future *in silico* models could potentially understand patient-specific parameters and, in turn, design personalized drug molecules.

While the discovery of patient-specific therapies would redefine the drug industry, researchers will first need to overcome several challenges. The most notable one is the problem of simulating a complex biological organism (Bender & Cortés-Ciriano, 2020). Simulating chemistry in isolation requires modeling a finite set of relatively well-understood parameters such as pressure or temperature (Bender & Cortés-Ciriano, 2020; Schneider et al., 2020). Such computational models already exist today. However, a substance predicted to be effective based on an assessment in a sterile, isolated environment may behave in unexpected ways in living organisms, which are much harder to describe using a computer model (Vamathevan et al., 2020). Pushing the envelope will require more advanced computational techniques and a better understanding of human biology (Bender & Cortés-Ciriano, 2020; Vamathevan et al., 2020). Technological leaps could also contribute to solving this problem; for example, quantum computers will enable researchers to simulate the world on an atomic level (Knight, 2018). As of today, however, the rapid development of patient-specific therapies remains only a future possibility.

**The transformation of the drug discovery landscape**

The AI-enabled pharmaceutical R&D market increased from US$200 million in 2016 to more than US$700 million in 2018 and is projected to keep growing (Deloitte, 2018). Tech giants like Google, Tencent, or Vodafone partner with pharmaceutical companies and, similarly to governments, are increasing their AI research investments (Deloitte, 2019a). The innovations discussed earlier are just the beginning of the transformation. Several studies hypothesize future revolutions such as automated drug discovery, rapid identification of generic alternatives to brand drugs, or personalized drugs (Paul et al., 2010; Schneider & Clark, 2019; Zhavoronkov et al., 2019b).

Most AI research requires access to large amounts of data, the availability of which is critical to advances in the field. While several large databases relevant for drug research are openly available, the amount of information they offer is minuscule compared to the size of proprietary databases siloed in pharmaceutical companies (Schneider et al., 2020). Traditionally, pharmaceutical companies' proprietary data was treated as a competitive advantage and guarded closely to the point where several reports about clinical trials became unavailable a few years after they were initially published (Hopkins et al., 2018). Fortunately, the landscape is changing, and several initiatives ensued to make use of the siloed data. Two similar projects were started by the European Innovative Medicines Initiative (IMI) and Massachusetts University of Technology (MIT). Both organizations partnered with pharmaceutical companies to analyze their proprietary data using ML methods (MELLODDY, 2019; MIT, n.d.). Conversely, British pharmaceutical company GlaxoSmithKline started sharing anonymized patient-level data from clinical trials in 2013 (GlaxoSmithKline, n.d. As data sharing initiatives become widespread, they will contribute to reducing the time required for discoveries.

**Discussion**

While most agree that the adoption of AI will impact the pharmaceutical industry, the nature of this impact remains debated. On the one hand, AI could end up being merely an incremental improvement to classic research methods. On the other, AI could become advanced enough to remedy the productivity crisis. Finally, AI could become the catalyst to a radical transformation of the entire industry. The rapidly growing market for pharmaceutical AI applications and the promising results of drug-enabled AI research supports the latter scenario. Classical research methods are increasingly more challenging to apply to the growing amount of available data: There are 4,500 known drug targets, $10^{60}$ possible drug compounds, 14,000 existing drugs, and the number of related results, studies, and databases increase each year (Reymond & Awale, 2012). Moreover, the bar for every new drug is higher than for its predecessors – a so called *Better than the Beatles* problem:

> Imagine how hard it would be to achieve commercial success with new pop songs if any new song had to be better than the Beatles, if the entire Beatles catalogue was available for free, and if people did not get bored with old Beatles records. We suggest something similar applies to the discovery and development of new drugs. Yesterday's blockbuster is today's generic. An ever-improving back catalogue of approved medicines increases the complexity of the development process for new drugs, and raises the evidential hurdles for approval, adoption and reimbursement. It deters R&D in some areas, crowds R&D activity into hard-to-treat diseases and reduces the economic value of as-yet-undiscovered drugs. (Scannel et al., 2012).

While AI advancements may lead to faster and more accurate research, the causes of the pharmaceutical productivity crisis go beyond companies' ability to process data. Other, perhaps more important, factors include lack of cooperation between key players, market incentives, and business strategies (Cockburn, 2016; DiMasi et al., 2016; Scannell et al., 2012). At first sight, AI-related considerations may seem orthogonal to these concerns. A closer inspection, however, reveals interconnectedness between all these aspects of the productivity crisis.

**AI and collaboration.**

Traditionally, the pharmaceutical industry's key players kept their key research a secret (Hopkins et al., 2018; Piller, 2020; Schneider et al., 2020). Some were motivated by gaining a competitive advantage over competitors, others lacked resources and expertise, and the rest simply never developed appropriate culture or policies (Barrett, 2020). The outcome has always been the same: New research projects could not benefit from prior knowledge (Beninger et al., 2017; Hayes & Hunter, 2012). As a result, different companies overlapped in efforts by working on similar drug targets, lead molecules, and safety assessments as others already did (Sanz et al., 2017; So et al., 2013). Some research projects were doomed to fail from the outset, and others required larger investments, more time, and more effort than they would have otherwise. The status quo contributed to the productivity crisis and, ultimately, disadvantaged everyone. Fortunately, in recent years, the secrecy culture has started to shift towards transparency and data sharing (Institute of Medicine, 2013). The rapid advancements in pharmaceutical AI played a significant role in inspiring and accelerating this shift.

The development of AI-enabled drug research methods depends on the availability of large volumes of data, yet most of the existing pharmaceutical data is proprietary (Schneider et al., 2020). In an ideal world, large players would benevolently join forces, pool their entire research together, and extend a combined effort to discover new breakthrough medicines. While such cooperation would likely significantly advance pharmaceutical research, in reality, it is unlikely to happen without facilitation or even legislation. Fortunately, the problem has been noticed by the governments, academia, tech companies, and even industry leaders. Several initiatives and partnerships ensued to accelerate AI development, including Pistoia, Allotrope, SALT, MELLODDY, MLPDS, eTOX, and ATOM (Allotrope, n.d.; Atom Science, 2017; eTOX Project, 2010; MELLODDY Project, 2019; MIT, n.d; Pistoia Alliance, 2020). Top companies like GlaxoSmithKline and Sanofi started sharing the results of their clinical trials, both successful and failed (Deloitte, 2019b; GlaxoSmithKline, n.d.; Sanofi, 2013). Several contractual partnerships between tech companies, pharmaceutical giants, and academia enabled AI research projects based on proprietary data (Deloitte, 2019b). For example, a company Atomwise established more than 60 such partnerships with pharmaceutical and academic entities, gaining access to resources that enabled the identification of a novel candidate drug molecule for multiple sclerosis (Atomwise, n.d.; Deloitte, 2019b). These promising developments are a step towards a more transparent industry, but only a step.

Ironically, in many data-sharing initiatives, large chunks of data donated by drug companies remained confidential in one way or another. For example, the eTOX consortium united several pharmaceutical companies who pooled their proprietary data in a central database governed by an entrusted third-party regulating access (eTOX project, 2010, 2018). A significant part of the database remained accessible only by the sharing party, allowing other consortium members to use it for training machine learning models but not to browse it directly. This partnership model proved to be successful – the eTOX consortium successfully built the largest database of molecules and several predictive models of their safety and toxicity (eTOXSys, n.d.; Sanz et al., 2017). The MELLODDY project had to overcome an even more challenging situation: Project participants agreed to share their confidential datasets under the condition that no data is extracted from their datacenters (IMI, 2020; MELLODDY, 2019). As training machine learning models require centralized access to the training data, the researchers had to push the envelope to overcome these limitations. IMI developed a novel method called "federated learning," which enables training statistical models using independent datasets while masking which outcomes were contributed by which dataset (2020). Projects like eTOX and MELLODDY prove that scientific progress does not have to be limited by market competition. More importantly, they show that even a small step towards data transparency goes a long way towards accelerating pharmaceutical AI.

**AI and market incentives.**

The competitive nature of the pharmaceutical sector is a source of several moral dilemmas. For example, the demand for rare diseases drugs is relatively small, providing little market incentives to invest in expensive research (Tambuyzer, 2010). It comes as no surprise that most known diseases are not treatable today despite fast-track procedures and other incentives aiming to encourage the research (Tambuyzer, 2010; VFA, 2015). Similarly, the cost of some life-saving drugs is too high for many to afford. Drug companies claim high R&D costs drive the price, yet a close analysis reveals that more money is spent on promoting existing drugs rather than researching new ones (Brekke & Straume, 2008). Furthermore, much of the research remains a trade secret, yet pooling the knowledge would significantly reduce the time and effort required to discover new drugs (Deloitte, 2019b). These dynamics may soon change as the increasing relevance of AI-driven methods shifts market incentives.

AI may diminish pharmaceutical patents' role – a vital component of industry business strategies today (Brekke & Straume, 2008; Deloitte, 2019b; KPMG, 2020; Open Markets Institute, n.d.). Patenting a drug grants the company exclusive rights for manufacturing and sales, leading to monopolies, high drug prices, and less innovation (Brekke & Straume, 2008; Open Markets Institute, n.d.). The monopolists can retain their strong position by merely exploiting the revenue from their existing products. Lowering the drug development barrier would threaten these large revenue streams, pushing the industry leaders to seek alternative revenue models. One such model would be improving existing drug therapies, for example, by increasing their safety, effectiveness, or offering them for a lower price. Another model would be exploring the vast space of the estimated 20,000 untreatable diseases by developing novel therapies. Finally, the companies could compete by further innovating the drug development process and inventing proprietary methods and models for faster and more accurate research (Deloitte, 2019b; KPMG, 2020; Narain et al., 2011). While each outcome would benefit society, pharmaceutical companies may fight to maintain their privileged position, as happened many times already (Open Markets Institute, n.d.). Some examples from recent history include price-fixing, paying competition to keep their drugs from the market (*pay–for–delay*), and preventing others from producing generic drugs by refusing to cooperate (Bartz, 2016, 2020; Federal Trade Commission, 2014). All in all, the market will likely change in the future, but the transition may not occur without friction.

The market will also change as tech companies, already setting their foot in the market, strengthen their position further. As consulting companies Deloitte (2019b) and KPMG (2020) report, tech giants are already active on the market, and pharmaceutical AI startup companies continue to raise record rounds of funding research. The mushrooming biopharma–tech partnerships and data-sharing initiatives will likely lift the largest growth barrier for tech companies: lack of data access. As tech companies continue to develop faster and more precise solutions, they may become capable of conducting large parts or even the entirety of the drug discovery process. There are three possible outcomes (KPMG, 2020): First, the AI companies may partner with pharmaceutical giants, for example, as outsourcing vendors. Second, AI companies may start developing drug therapies on their own, becoming a new force on the pharmaceutical market and threatening today's leaders. Third, tech companies could become partners to some and threat to others. In either scenario, they will be a force to be reckoned with.

Many other possible AI-related disruptions could transform the market. For example, China introduced a fast track for drugs targeting rare medical conditions to encourage new research projects (Deloitte, 2019b). General-purpose drugs targeting specific proteins could give place to personalized medicine tailored to each patient's unique characteristics (Deloitte, 2019a, 2019b; KPMG, 2020). Data-sharing initiatives could lead to sharing datasets and algorithms publicly, enabling crowdfunded open-source and non-profit drug research projects independent from pharmaceutical companies (Deloitte, 2019a, 2019b). Still, the future remains unknown, and the transformation may not happen as quickly or be as radical as some predict. That considered, the partnerships and regulatory changes already happening today will have a ripple effect in the years to come.

**AI and safety and efficacy of drugs.**

One of the research directions in pharmaceutical AI is predicting the adverse effects of promising drug molecules early in the process. Despite extensive clinical trials, toxicity and side-effects remain the leading cause of withdrawing drugs from the market (Siramshetty et al., 2016). The ability to predicting these characteristics upfront would significantly increase the safety of new medicines. In recent years, the eTOX project provided significant advancements by building the largest molecule toxicity database and using it to build more than 200 predictive models (eTOX, 2010, 2018; Sanz et al., 2017). As there is still much space to innovate, several other initiatives and pharmaceutical AI startups are currently researching the same area. Some studies went beyond toxicity and explored predicting the entire research project's future success early on (Vamathevan et al., 2020). A historical analysis of drug research projects revealed a few potential early predictors correlated with the later success of researched drug molecules success (Vamathevan et al., 2020). While considering such results, it is essential to remember that a correlation indicates past performance, not future results. Pharmaceutical research keeps changing as new studies, modalities, and law regulations emerge (Schneider et al., 2020; Vamathevan et al., 2020). The most easily discoverable drugs may have been already found (Cockburn, 2016; Scannel et al., 2012). Successful research projects conducted in the past were different from those conducted today, and those of the future will differ even more (Schneider et al., 2020; Vamathevan et al., 2020). Regardless of whether the drug compound's success may be predicted upfront or whether a computer-based solution may predict toxicity with similar accuracy levels as animal and human trials, the data-driven approach will still disrupt drug safety. The partnerships and data-sharing initiatives ensuing in the process have an impact by themselves, enabling researchers to access much richer datasets than ever before.

Another AI frontier is repurposing existing drugs: A process of identifying different diseases sharing at least some drug targets with the ones already treatable (Hirogani et al., 2019). The inherent advantage of repurposing is the availability of accurate data. For one, existing drugs have already undergone clinical trials; For another, the market is a rich source of data about adverse effects and unanticipated interactions with other medications. What makes repurposing possible is the high likelihood of druggable targets being shared between many diseases (Hirogani et al., 2019). Several biopharma companies already started investigating their proprietary molecules in the search for new applications. As with many other research avenues, the most considerable challenge remains the availability of data. To know which targets are shared among investigated diseases, researchers will first need to accurately determine the exact role of different targets in each condition.

The impact of AI on drug safety goes beyond the research phase. Several AI-enabled manufacturing techniques are already in use today, following FDA's recommendation to adopt the "Quality by Design" approach in drug manufacturing (Aksu et al., 2013; Gams, 2014). Novel techniques were also developed for Quality Assurance and Quality Control to aid human operators, leading to reduced workload and error rate. These applications already increased the safety and consistency of manufactured medicine. As the research continues, more AI methods will be integrated into manufacturing, leading to even less recall and safe and consistent dosages.

**Timelines for the AI revolution.**

Pharmaceutical AI companies are growing at an extraordinary pace thanks to unprecedented investments from governments, venture capital companies, and tech giants worldwide (Deloitte, 2018, 2019a, 2019b). In the third quarter of 2020, these investments amounted to US $2 billion in the United States alone, two times more than in the second quarter (CBInsights, 2020a, 2020b). The number of drug-discovery focused AI startups has increased from 89 in 2014 to 217 in 2020 (BiopharmaTrend, 2020; Williams, 2020). The enthusiasm is well-founded as tech companies ship tangible AI-based solutions every year. For example, IBM launched an AI-based platform in collaboration with Pfizer to aid immune-oncology research and clinical trials design (IBM, 2016). Moreover, Microsoft launched *Project Hanover*, a natural language processing tool aiding in personalizing cancer drugs by identifying information related to patients' specific profile in the medical literature (Microsoft, 2019). Furthermore, Cyclica, a startup from Toronto, partnered with industry leaders and launched AI-based platforms for optimizing candidate drug molecules and *de novo* design of new compounds against requested targets (Deloitte, 2019b). All in all, tech companies attracted large investments and are now expected to ship innovative solutions rapidly.

The most optimistic reports predict that the sector will grow exponentially, and by 2030 tech players will perform most of the R&D process and even register drugs by themselves (Deloitte, 2019a, 2019b; KPMG, 2020). These reports reason that the increase in input (data and investments) will lead to a corresponding increase in output (innovation). In such a scenario, today's promising early research would shortly evolve into the next generation of AI tools and platforms, enabling drug research projects to reach clinical trials in a matter of weeks or months instead of years. Those AI drug discovery companies that remain independent and develop their own products would become key to pharmaceutical research (Deloitte, 2019b; KPMG, 2020). Today's industry leaders would be forced to either rent software and expertise out from AI vendors or pool their resources together to overcome high R&D costs. Given the broader context, such a development is not unlikely in the longer term.

While the industry is headed towards radical transformation, the journey will likely require more time than just a decade (Bender & Cortés-Ciriano, 2020). So far, many AI products have provided only an incremental improvement rather than a disruption, while others need further refinements before they could be widely adopted (Schneider & Clark, 2020; Schneider et al., 2020). While increased funding will likely lead to more AI advancements, the productivity crisis in pharmaceutical R&D illustrates that innovation does not linearly follow the number of dollars invested (Bender & Cortés-Ciriano, 2020; Scannel et al., 2012). Only a handful of pharmaceutical AI studies available today have practical applications (Schneider et al., 2020). The hype around recent AI developments makes it easy to overestimate their actual impact.

The AI revolution may not take off until pharmaceutical AI companies refocus their research on improving decisions' quality. Bender and Cortés-Ciriano (2020) studied the impact of improving either speed, failure rate, or cost at any other drug research phase on the project's cost and the quality of the outcomes. They found that even though a reduction in clinical trials' failure rate would have the most considerable impact by a large margin, much of the pharmaceutical AI research is centered around improving the speed and cost of various steps. The study reports that a consequence of this approach is the prevalence of high-throughput methods of identifying and validating drug targets and molecules. Such methods predict properties like drug molecule toxicity, 3D structure, or on-target activity (Bender & Cortés-Ciriano, 2020; Deloitte, 2019b). While these metrics are intuitively sensible, they are generally poor predictors of clinical outcomes as they fail to predict nuanced effects a drug may have in biological organisms (Bender & Cortés-Ciriano, 2020; Vamathevan et al., 2019). In contrast, today's low-throughput techniques provide better clinical results predictability, although they are not practically applicable on a larger scale (Bender & Cortés-Ciriano, 2020; Schneider et al., 2020). In the future, researchers may invent methods connecting the speed of today's high-throughput techniques with the accuracy of the low-throughput methods. Today, shifting the focus of pharmaceutical AI research towards more qualitative methods may be pivotal to the transformation.

## Conclusions

The emerging AI revolution will help the pharmaceutical industry move from productivity crisis to peak productivity, enabling a surge of safe and affordable medicines. Such a change will not happen overnight, but the industry is already moving in that direction today. The pharmaceutical AI already boosted a handful of drug research projects and attracted record investments, despite still being in infancy. The largest obstacle remains the availability of the data and the computational complexity of models. Overcoming these limitations will lead to the discovery of more sophisticated techniques allowing for the development of effective therapies in a fraction of the time and budget required today.

However, even if the research does not yield revolutionary computational methods, the initiatives supporting pharmaceutical AI advancements will have a long-lasting disruptive effect. The data-sharing alliances and partnerships will foster collaboration, reduce knowledge barriers, and lead to discoveries and better decisions. The tech companies will continue gaining experience, establishing data-sharing alliances, and will stay on the market as vendors and catalysts for future innovation. Furthermore, today's AI has already enabled innovations like crowdsourced research techniques and automated chemical laboratories (Deloitte, 2019b; KPMG, 2020). The industry is already being transformed, and innovative future initiatives will only magnify this change. Whether or not revolutionary drug discovery methods ever emerge from the AI wave, something more meaningful and disruptive already did – a chance for open science.